\documentclass[twocolumn,           
               showpacs,            
               preprintnumbers,     
               aps,                 
               prd,                 
               a4paper,             
               groupedaddress,  
               nofootinbib,         
               tightenlines,        
               floats,floatfix      
               ]{revtex4}

\usepackage{graphicx}
\usepackage{subfigure}

\usepackage{latexsym}
\usepackage{amsmath,amssymb}        
\usepackage[draft=false]{hyperref}
\usepackage{threeparttable}

\input colordvi

\begin{document}



\title{On the possibility of braneworld quintessential inflation}
\author{Mafalda Dias}
\affiliation{Astronomy Centre, University of Sussex, Brighton BN1 9QH,
  United Kingdom}
\author{Andrew R.~Liddle}
\affiliation{Astronomy Centre, University of Sussex, Brighton BN1 9QH,
  United Kingdom}
\date{\today}


\begin{abstract}
We examine the possibility of achieving quintessential inflation,
where the same field serves as both inflaton and quintessence, in the
context of a five-dimensional braneworld. Braneworld cosmology provides an
appropriate environment as it permits inflation with much steeper
potentials than the conventional scenario, which is favourable to a late-time quintessence. We explore a wide space of models, together with contemporary observational data, to determine in which contexts such a picture is possible. We find that such a scenario, although attractive, is in fact impossible to achieve for the potentials studied due to the restrictiveness of current data.

\end{abstract}

\maketitle


\section{Introduction}

In the past decade there has been a lot of interest in the study of extra dimensions, in particular the scenario where our apparent Universe lives in a  three-dimensional hypersurface, a 3-\textit{brane}, embedded in a space with more dimensions (for a review see Ref.~\cite{quevedo}). This is the so-called \textit{braneworld} picture, which brings new possibilities and perspectives to cosmology. In this work we will consider the simplest case of a braneworld scenario, where there is only one extra dimension, and analyze the impact of this modification in the attempt of getting quintessential inflation \cite{PV}. 

Our framework is of Kaluza--Klein type, but where, instead of compactifying the extra dimension, one assumes that matter fields are confined to the brane and that gravity is allowed to act in the five-dimensional embedding space, generally called the \textit{bulk}.  We adopt the simplest such set-up, known as Randall--Sundrum Type II \cite{RS2}.
By solving the Einstein equations in the vicinity of the brane,  one finds a modified Friedmann equation that determines the cosmological evolution caused by fields confined to our 3-brane, such as a scalar field \cite{langlois}.
In this case, the inclusion of an extra dimension in the theory introduces an extra friction term in the high-energy limit, which allows inflation to occur with much steeper potentials than usual \cite{maartens,steep}. This is favourable to get a second inflation today out of the same scalar field, \textit{i.e.}\ a quintessence, from the behaviour that such potentials have at late times. 

At lower energies, the braneworld-induced friction term is negligible and the scalar field spends a period completely dominated by its kinetic energy. In this era, usually called \textit{kination}, the field density decreases very rapidly, $\rho \propto a^{-6}$ where $a$ is the scale factor, permitting unusual reheating processes. The examples that will be explored later are reheating by gravitational particle production \cite{gravprod}, where radiation-like particles created during inflation can come to dominate the Universe during kination, and curvaton reheating where energy density is stored in a second scalar field to be later released by its decay. 

The existence of a kination together with the possibility of having steep potentials makes this scenario appealing for the implementation of quintessential inflation \cite{PV,steep}. With a suitable choice of scalar field potential and reheating process one can try to fit observational data from inflation, big bang nucleosynthesis (BBN) and quintessence. There have been several studies of such models \cite{steep, huey,sahni,NC,curvaton, portugas}, but none of them satisfy all the constraints brought by contemporary data. In this present work, we make a thorough review as well as expanding the space of models to fully understand to what extent it is possible to get inflation and quintessence out of the same scalar field in this simplified view of the braneworld paradigm. We will see that for the potentials analyzed, quintessential inflation cannot be achieved. 

In this paper, we discuss potentials with a simple structure, making only brief comments in Section IV on more complex ones, \textit{i.e.} potentials that combine different forms at early and late times (for example Refs.~\cite{PV,portugas}). Of course, we want potentials with good late-time behaviours and, as will be shown, the exponential and inverse power-law types are in that class. Therefore these are the two potentials analyzed by us in detail.

The paper is organized as follows. In Section II we specify the models that we will test, presenting separately what is happening at early and late times. In Section III we impose constraints on these models based on their viability in comparison with observations. Again, the analyses of the constraints for the inflation period and late-time quintessence are treated independently. In Section IV we combine the constraints from both periods, to study the viability of quintessential inflation. Finally, in Section V we summarize the main results we obtained.

\section{Braneworld quintessential inflation models}

The modified Friedmann equation for a five-dimentional braneworld has the form \cite{langlois}
\begin{eqnarray}
H^{2}=\frac{8\pi}{3 m_{4}^{2}}\rho \left[1+ \frac{\rho}{2\lambda}\right] + \frac{\Lambda_{4}}{3}+ \frac{\epsilon}{a^{4}}
\end{eqnarray} 
where $H$ is the Hubble parameter, $\Lambda_{4}$ is a cosmological constant, $m_{4}$ is the four-dimensional Plank mass and $\lambda$ is the brane tension. The last term is related to the effect of bulk gravitons on the brane, but vanishes very quickly as inflation starts (spatial curvature is ignored for the same reason). Since we are interested in quintessence models, we assume that the four-dimensional cosmological constant is zero.

The Friedmann equation then takes the simple form
\begin{eqnarray}
H^{2}=\frac{8\pi}{3 m_{4}^{2}}\rho \left[1+ \frac{\rho}{2\lambda}\right] 
\end{eqnarray} 
where the second term will give an extra friction to the scalar-field evolution. The scalar field $\phi$ that we will consider to give rise to inflation and quintessence is confined to the brane, following the standard evolution:
\begin{eqnarray}
\ddot{\phi}+3H\dot{\phi} = -\frac{dV}{d\phi}.
\end{eqnarray}

\subsection{Inflation}

The simplest model of this class, presented in Ref.~\cite{steep}, generates inflation with an exponential potential of the general form:\footnote{Note that many papers define $\alpha$ to be smaller by a factor $\sqrt{8\pi}$, by using the reduced Planck mass in the definition.}
\begin{equation}
V=V_{0}\exp(-\alpha\phi/m_4).
\end{equation}
Here $\alpha$ is taken to be greater than $\sqrt{16\pi}$, a regime in which the potential is too steep to sustain inflation in Einstein gravity.
In the high-energy limit, with $\rho \gg \lambda$, the stronger friction term ensures that inflation can occur. During inflation, due to quantum fluctuations, radiation-like particles are created, in the generally-called gravitational particle production process \cite{gravprod}.
At a low-energy limit, when $\rho$ becomes smaller than $\lambda$, the extra term of the Friedmann equation becomes unimportant and the field starts evolving as standard in a steep potential. Inflation ends and the field, experiencing such a steep potential, becomes dominated by its kinetic energy. During this kination, the effect of the potential is negligible and the energy density of the field decreases very rapidly as  $\rho \propto a^{-6}$. This abrupt loss of energy allows the gravitationally-produced radiation to dominate before BBN. This reheats the Universe without requiring our scalar field to decay.

The predictions of this model for the inflationary parameters $n_{\rm s}$, the spectral index of scalar perturbations, and $r$, the ratio of tensor to scalar perturbations, can be easily obtained using the modified slow-roll parameters of Ref.~\cite{maartens}, and have the simple form \cite{steep}
\begin{eqnarray}
n_{\rm s} & =& 1-\frac{4}{N+1} \\
r & =& \frac{24}{N+1}
\end{eqnarray}
where $N$ is the number of e-foldings. It has been shown that in this case $N$ can be unambiguously calculated to be 70 \cite{sahni}. This results in the values $n_{\rm s}=0.944$ and $r=0.33$, which in combination are clearly excluded by current observations \cite{Komatsu5yrWMAP, Komatsu7yrWMAP}.

This is one of the problems of this model. Another is related to a relic gravitational waves imprint. Gravitational waves can be treated as a massless radiation field that decays as $\rho_{\rm GW}\propto a^{-4}$ when it is in the subhorizon regime. Short-scale gravitational waves, created by the same gravitational production mechanism, enter the subhorizon regime right after inflation, during the kinetic regime where the dominating background evolves as $\rho \propto a^{-6}$. This means that these gravitational waves will have a boost in their relative energy density $\propto a^{2}$. Such a behaviour leaves an imprint that destroys BBN, excluding the model \cite{sahni}.

A solution to both these problems can come from the choice of another type of reheating, which brings the values of $|n_{\rm s}-1|$ and $r$ down as well as limiting the duration of the dominating kinetic period. This can be achieved with the inclusion of a curvaton field \cite{wands,curv2} whose decays reheat the Universe \cite{curvreh,curvaton}.

The curvaton is another field which coexists with the inflaton, that is subdominant and massless during inflation, and which can be responsible for the primordial curvature perturbations. For simplicity, we consider that this field has a quadratic potential of the form:
\begin{equation}
U(\sigma)=\frac{1}{2} m^{2}\sigma^{2}
\end{equation}
where $m$ is the mass of the curvaton. To ensure this is effectively massless during inflation we impose the condition $m \ll H_{f}$, the Hubble parameter at the end of inflation. Under these conditions, the value of the curvaton field, $\sigma$, remains constant during inflation \cite{wands,curvaton}, apart from acquiring quantum-originated fluctuations from the expanding background.

If we choose $m$ correctly, during the kinetic period the field becomes effectively massive, \textit{i.e.}\ $m \approx H$. This should happen at an early time to avoid the short-scale gravitational waves problem. The field then starts to oscillate, which corresponds to behaviour as matter, and its energy density falls as $\rho_{\sigma} \propto a^{-3}$ \cite{wands}. It then becomes dominant quite rapidly. 

The choice of $m$ and of $\sigma_i$, the value of the field at the start of inflation, needs to be made in such a way that inflation and BBN can occur successfully. These constraints will be analyzed in detail in Section III.A. 

We will assume that the curvaton decays into radiation by some process with decay rate $\Gamma$, when $\Gamma = H$. The field could start its decay while the kinetically-driven inflaton is still dominating, which can bring some calculational complications \cite{wands, curvaton}. In this work, for simplicity, we will assume that the curvaton only starts decaying after being dominant. 

How are primordial perturbations created in this scenario? During inflation, the curvaton experiences quantum fluctuations, like the inflaton. These fluctuations of $\sigma$ freeze at horizon crossing and have a spectrum 
\begin{equation}
P_{\sigma}=\left(\frac{H_{*}}{2\pi}\right)^{2}
\end{equation}
where the star subscript denotes horizon crossing \cite{wands}. Then, after inflation, when the curvaton becomes massive and oscillating, these fluctuations are converted into scalar curvature perturbations. The spectrum of these curvature perturbations has the form \cite{wands}
\begin{equation}
P_{\zeta}=\frac{1}{9\pi^{2}}\frac{H_{*}^{2}}{\sigma_{i}^{2}}.
\label{Hi}
\end{equation}
The scale dependence of $P_{\zeta}$ is the same as that of $P_{\sigma}$, and so the spectral index has the simple form
\begin{equation}
n_{\rm s}=1-2\epsilon,
\end{equation}
where $\epsilon$ is the slow-roll parameter. This matches the spectral index of primordial tensors and is usually different from that of the inflaton-induced perturbations.

The parameter $r$ (we follow the convention as in Ref.~\cite{anthony}), defined as 
\begin{equation} 
r=16\frac{h_{\rm GW}^{2}}{P_{\zeta}^{2}},
\end{equation}
where $h_{\rm GW}$ is the spectrum of tensor perturbations, also suffers alterations. Tensor perturbations in the metric are produced during inflation; they have the general form of \cite{GW}:
\begin{equation}
h_{\rm GW}^{2}=\frac{H_{*}^{2}}{\pi m_{4}^2} F^{2},
\label{hGW}
\end{equation}
where $F$ is a braneworld correction that in the high-energy limit takes the form
\begin{equation}
F^{2}=\frac{3\sqrt{3}H_{*}m_{4}}{4\sqrt{\pi\lambda}}.
\end{equation}

In general, the scalar curvature perturbations $P_{\zeta}$ should be a sum of the perturbation of the inflaton and curvaton. In our case, we are assuming that the curvaton perturbations are dominant, so $r$ has the form
\begin{equation}
r=\frac{72\pi\sigma_i^{2}}{m_{4}^{2}}F^{2}.
\end{equation}
In Section III.A, where constraints are applied to the values of $\sigma$ and $m$, we will show how possible is the inclusion of a successful curvaton to the model, and if so how efficient it is in matching the WMAP observations.
\\

The other simple potential that we analyzed in our study is the inverse power law:
\begin{equation}
V=V_{0}\left(\frac{\phi}{m_{4}}\right)^{-\alpha}.
\end{equation}
This potential allows an arbitrary amount of early-Universe inflation for any $\alpha>2$, as can be seen by the modified expression for the number of $e$-foldings in Ref.~\cite{maartens}.

Just as in the case of the exponential potential, we could seek a good model by reheating the Universe with gravitational particle production. But the kinetic dominated period would still be too long, so the short-scale gravitational waves would remain a problem.

\begin{figure}[t]
\includegraphics[width=0.9\linewidth]{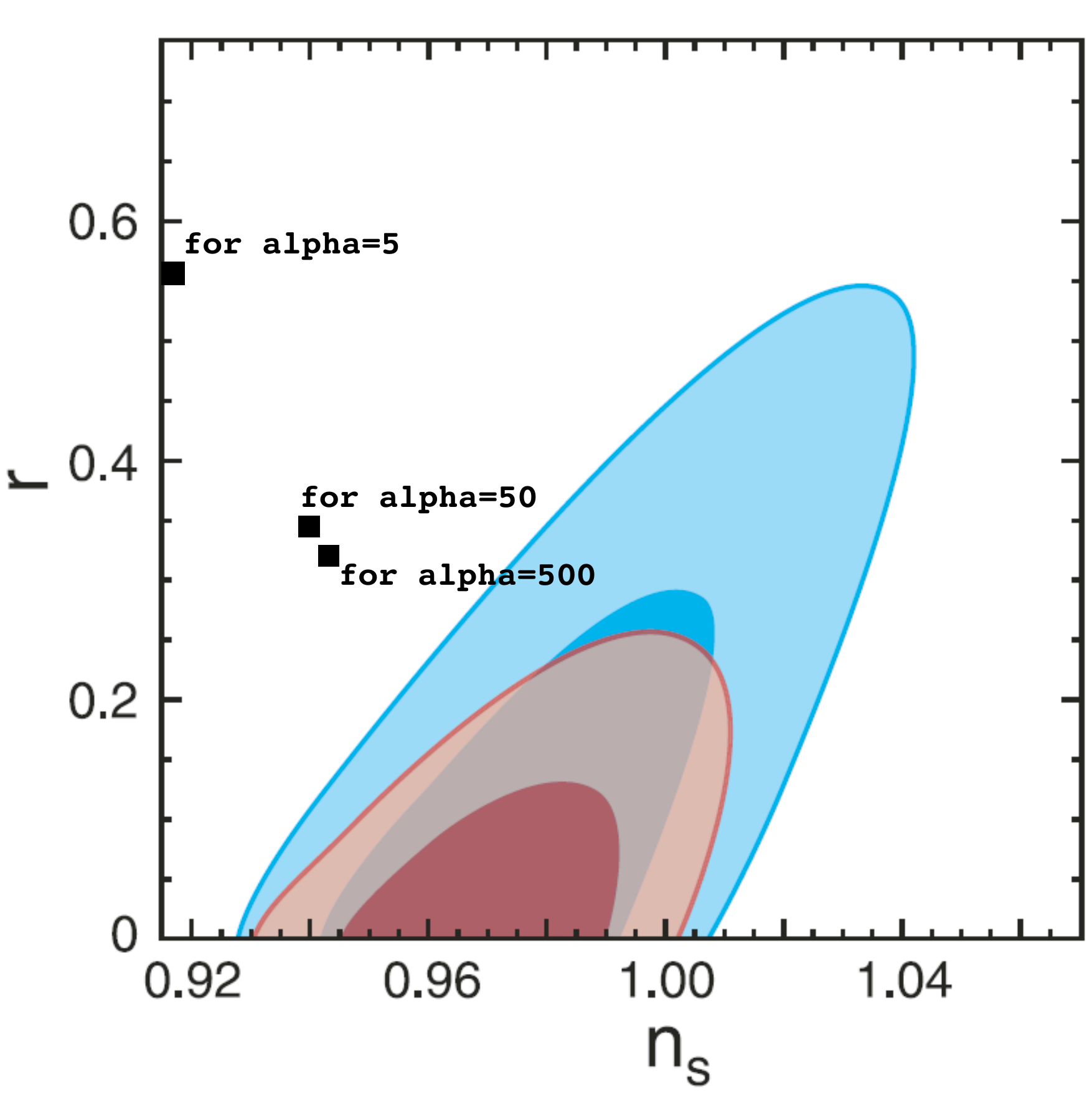}
\caption{Inverse power-law potentials without curvaton. Comparison of the values of $n_{\rm s}$ and $r$ with observations from WMAP five-year data, for different values of $\alpha$. The exponential potential would lie under the $\alpha=500$ point. The plot shows the $68\%$ and $95\%$ confidence limits. The outer limits correspond to WMAP data alone, and the inner to WMAP data combined with supernovae and baryonic acoustic oscillation data. We can clearly see that this model is excluded by observations. Adapted from Ref.~\cite{Komatsu5yrWMAP}.}
\label{wmapinvnoc}
\end{figure}

Further, again using the modified slow-roll parameters \cite{maartens}, we can calculate $n_{\rm s}$ and $r$:
\begin{eqnarray}
n_{\rm s}& =& 1-\frac{4\alpha -2}{N(\alpha-2)+\alpha} \\
r & = & \frac{24\alpha}{N(\alpha-2)+\alpha}.
\end{eqnarray}
As can be seen in Fig.~\ref{wmapinvnoc}, these values are again excluded by WMAP.

So, as in the case of the exponential potential, to try to get a successful inflation, we need to include a curvaton in the model. In Section III.A we will show constraints on $\sigma_{i}$ and $m$ for this specific potential, to try to understand if a curvaton can in fact help the creation of inflation with good observational fit.

 \subsection{Quintessence}	
 
In the previous subsection, we sought to create a model that could give rise to successful inflation. Now it is necessary to check if these models can give rise to quintessence, and if so, under which conditions. To do so, we need to analyze the late-time behaviours of the potentials studied.

Kination won't last forever. The strong decrease of the energy density makes the kinetic energy decrease to a point where the potential is no longer negligible. The behaviours of such scalar fields that have been subdominant for a certain period of time is well known \cite{scaling,SWZ}. 

Particularly interesting solutions when  $\rho_{\phi} < \rho_{\rm back}$ have the general name of scaling and tracking solutions, since in these cases the scalar field will `follow' the evolution of the dominating background fluid. These two types of solution differ in the way the scalar field evolves. If we consider that the background fluid evolves as $\rho_{\rm back} \propto a^{-n}$ and that the scalar field evolves as $\rho_{\phi} \propto a^{-m}$, scaling solutions will be the ones where $n=m$ and tracking the ones where $n \neq m$. It is easy to see that such evolution is interesting, especially if we could get the scalar field tracking radiation and matter and only start dominating today.
As shown in Ref.~\cite{scaling}, the only potentials that give rise to the above tracking  behaviours are exponentials, inverse power laws and some positive power laws, although the last option is not interesting for connecting with inflation. Furthermore, it was also shown that these solutions are attractors.

Scalar fields with exponential potentials have scaling solutions at late times, with $n=m$. Since the ratio between dominant fluid and scalar field is always a constant, there is no possibility for late-time quintessence if we are in the scaling regime.

Scalar fields with inverse power laws have tracking solutions where $n<m$. This means that the scalar field energy density is increasing with time compared to the one of the dominant fluid. This is exactly the right condition to try to get quintessence. When the scalar field density approaches the one of the background fluid, the assumption that $\rho_{\phi} < \rho_{\rm back}$ breaks down, and the field starts dominating, until its density is $\Omega_{\phi} \approx 1$ and its equation of state is $w \approx -1$. In Section III.B we will analyze whether, beside being able to produce acceleration at late times, this model can also satisfy the constraints on such acceleration today, \textit{i.e.} $\Omega \approx 0.75$ and $w \lesssim -0.8$ \cite{Komatsu5yrWMAP, Komatsu7yrWMAP}.\\

An alternative regime occurs when the scalar field kination period lasts for so long that the field overshoots its tracking solution and then freezes due to redshifting of the kinetic term, later approaching that solution from below. If the overshoot is so great that the field only approaches the tracking solution around the epoch where tracking solution would be coming to dominate the fluid, we can enter directly into quintessence domination without a tracking period. This regime is referred to as creeping or thawing \cite{thawing} quintessence, and was first explored in our present context in Ref.~\cite{huey}. The energy density is initially approximately constant in this circumstance and the field behaves as a cosmological constant with $w \approx -1$. When the energy density of the scalar field gets close enough to the background energy density, the field begins to evolve to $w>-1$. 

With an inverse power-law, once thawing acceleration begins it will typically persevere. By contrast, the late-time solution for a steep exponential potential is a non-accelerating scaling solution, and in that case the field can only temporarily drive acceleration \cite{CLW}.

In Section III.B the above possibilities will be explored to see how well they can reproduce today's conditions on quintessence.

\section{Constraints}

\subsection{Inflation}

In this subsection we present the constraints that need to be imposed on the scalar field plus curvaton scenario, so that inflation occurs in a successful way. In fact, this will correspond to constraints on the space of values allowed for $\sigma_{i}$ and $m$, leaving as free parameters only $N$, the number of $e$-foldings, and $\alpha$, the potential parameter. 

Before presenting the constraints, let us consider some useful relations. In the high-energy limit, \begin{equation}
H \propto V
\end{equation}
and $V_{(i)}/V_{(e)}$, where $(i)$ and $(e)$ refer to the beginning and end of inflation, can be expressed as a function of $N$, by using the modified expression for $N$ given in Ref.~\cite{maartens}. For exponential potentials:
\begin{equation}
\frac{V_{(i)}}{V_{(e)}}=N+1
\label{ViVeexp}
\end{equation}
and for inverse power-law potentials:
\begin{equation}
\frac{V_{(i)}\phi_{(i)}^{2}}{V_{(e)}\phi_{(e)}^{2}}=N+1 \iff \frac{V_{(i)}}{V_{(e)}}=(N+1)^{\alpha/(\alpha-2)}.
\label{ViVeinv}
\end{equation}

Using the modified expression for the slow-roll parameter $\epsilon$, and setting it to unity, we can get an expression for $V_{e}$. For exponential potentials we obtain
\begin{equation}
V_{(e)}=\frac{\alpha^{2} \lambda}{4\pi}
\label{Vendexp}
\end{equation}
and for inverse power-law potentials:
\begin{equation}
V_{(e)}=\frac{m_{4}^{2}}{4\pi}\alpha^{2} \lambda \phi_{(e)}^{-2}.
\end{equation}

We don't so far know the value of $\phi_{(e)}$ that shows up in this expression. To obtain it, we performed a numerical calculation of the equations of motion of the system during inflation. As a result, we obtained an approximate expression for $V_{(e)}$ depending only on $V_0$, $\alpha$ and $\lambda$:
\begin{equation}
V_{(e)}=0.06\lambda^{\alpha/(\alpha-2)}V_{0}^{-2/(\alpha-2)}\alpha^{2}.
\label{Vendinv}
\end{equation}
In the following calculations we will assume this expression for $V_{(e)}$.  For simplicity, we also assume that kination starts immediately after inflation. 

Now we present the five conditions that must be satisfied for successful inflation:\\

1. The curvaton needs to be subdominant when it becomes massive, to avoid a curvaton-driven inflation. This condition is expressed as:
\begin{equation}
\left.\frac{\rho_{\sigma}}{\rho_{\phi}}\right|_{(m)} \ll 1
\label{condition1basic}
\end{equation}
where the subscript $(m)$ refers to the period when the curvaton becomes massive. 
During kination, the Hubble parameter evolves as $H \propto a^{-3}$, so
\begin{equation}
\frac{m}{H_{\rm kin}}=\frac{a^3_{\rm kin}}{a^{3}_{m}},
\end{equation}
and the Friedmann equation can be simplified to
\begin{equation}
H^{2}=\frac{8 \pi}{3 m_{4}^{2}}   \rho_{\phi}.
\end{equation}
We can then work out condition (\ref{condition1basic}):
\begin{equation}
\left.\frac{\rho_{\sigma}}{\rho_{\phi}}\right|_{(m)} = \frac{m^{2}\sigma_{i}^{2}}{2\rho_{\phi}^{(k)}\left(a_{(k)}/a_{(m)}\right)^6} = \frac{\sigma^{2}_{i} 8\pi}{6m_{4}^{2}} \ll 1
\end{equation}
where we use $(k)$ to refer to the beginning of kination.
The condition is then simply:
\begin{equation}
\frac{\sigma_{i}^{2}}{m_{4}^{2}} \ll \frac{3}{4\pi}.
\end{equation}

2. The curvaton needs to be subdominant at the end of inflation as well, to avoid a curvaton-driven inflation. This can be expressed as
\begin{equation}
\frac{U_{(e)}}{V_{(e)}} = \frac{m^{2}\sigma_{i}^{2}}{2V_{(e)}}\ll 1.
\end{equation}

3. The decay of the curvaton needs to occur before nucleosynthesis, so that this is successful, and after curvaton domination. We have the conditions:
\begin{equation}
H_{\rm nucleo}=10^{-40}m_{4} < \Gamma <H_{\rm (eq)}
\label{condition3basic}
\end{equation}
where the subscript `(eq)' refers to the curvaton--scalar field equivalence epoch.
We can write
\begin{equation}
1=\left.\frac{\rho_{\sigma}}{\rho_{\phi}}\right|_{\rm (eq)}=\frac{m^{2}\sigma_{i}^{2}a_{(m)}^{3}a_{\rm (eq)}^{6}}{2\rho_{\phi}^{(k)}a_{\rm (eq)}^{3}a_{k}^{6}}=\frac{4\pi m^{2} \sigma_{i}^{2}a_{(m)}^{3}a_{\rm (eq)}^{3}}{3 m_{4}^{2}H_{(k)}^{2}a_{(k)}^{3}a_{(k)}^{3}}
\label{equivalence}
\end{equation}
which is equivalent to having
\begin{equation}
H_{{\rm eq}}=\frac{4\pi \sigma_{i}^{2}m}{3m_{4}^{2}}.
\end{equation}
So condition (\ref{condition3basic}) is just
\begin{equation}
10^{-40} m_{4} \ll  \frac{4\pi \sigma_{i}^{2}m}{3m_{4}^{2}}.
\end{equation}

4. Inflation perturbations cannot be important since we wish to get perturbations entirely from the curvaton. Without the presence of the curvaton, we could estimate the value of $\lambda_{\rm no\  \sigma}$ from the COBE normalization of the density perturbations originated by inflation. To guarantee that curvaton-induced perturbations are much bigger than the ones originated by the inflaton, we just need to ensure that
\begin{equation}
\lambda \ll \lambda_{{\rm no}\  \sigma}.
\end{equation}

5. The last condition that needs to be imposed is that short-scale gravitational waves are within limits. We can express this as
\begin{equation}
\left.\frac{\rho_{\rm GW}}{\rho_{\sigma}}\right|_{\rm (eq)} = \left.\frac{\rho_{\rm GW}}{\rho_{\phi}}\right|_{\rm (eq)} \ll 1.
\end{equation}
Using the expression for the evolution of energy density of gravitational waves during the kinetic period from Ref.~\cite{curvaton},
\begin{equation}
\rho_{\rm GW}=\frac{32}{3\pi} h^{2}_{\rm GW} \rho_{\phi}\left(\frac{a}{a_{(k)}}\right)^{2},
\label{rhoGW}
\end{equation}
plus the amplitude of primordial gravitational waves, Eq.~(\ref{hGW}) which can be rewritten as
\begin{equation}
h^{2}_{\rm GW}=\frac{2V_{(i)}^{3}}{\lambda^{2}m_{4}^{4}},
\label{hGW1}
\end{equation}
we can estimate our condition.\\

\begin{figure*}[t]
\includegraphics[width=0.32\linewidth]{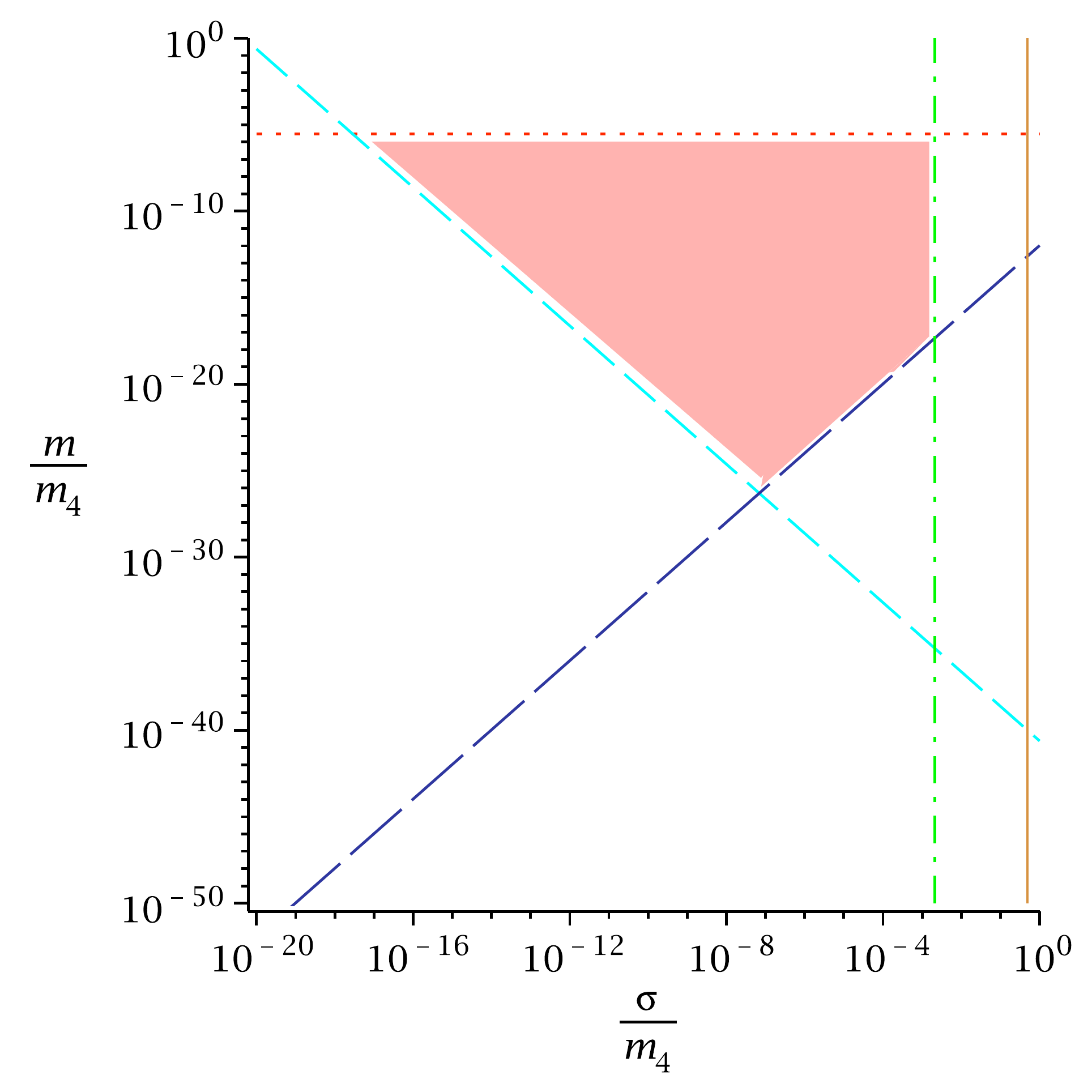}
\includegraphics[width=0.32\linewidth]{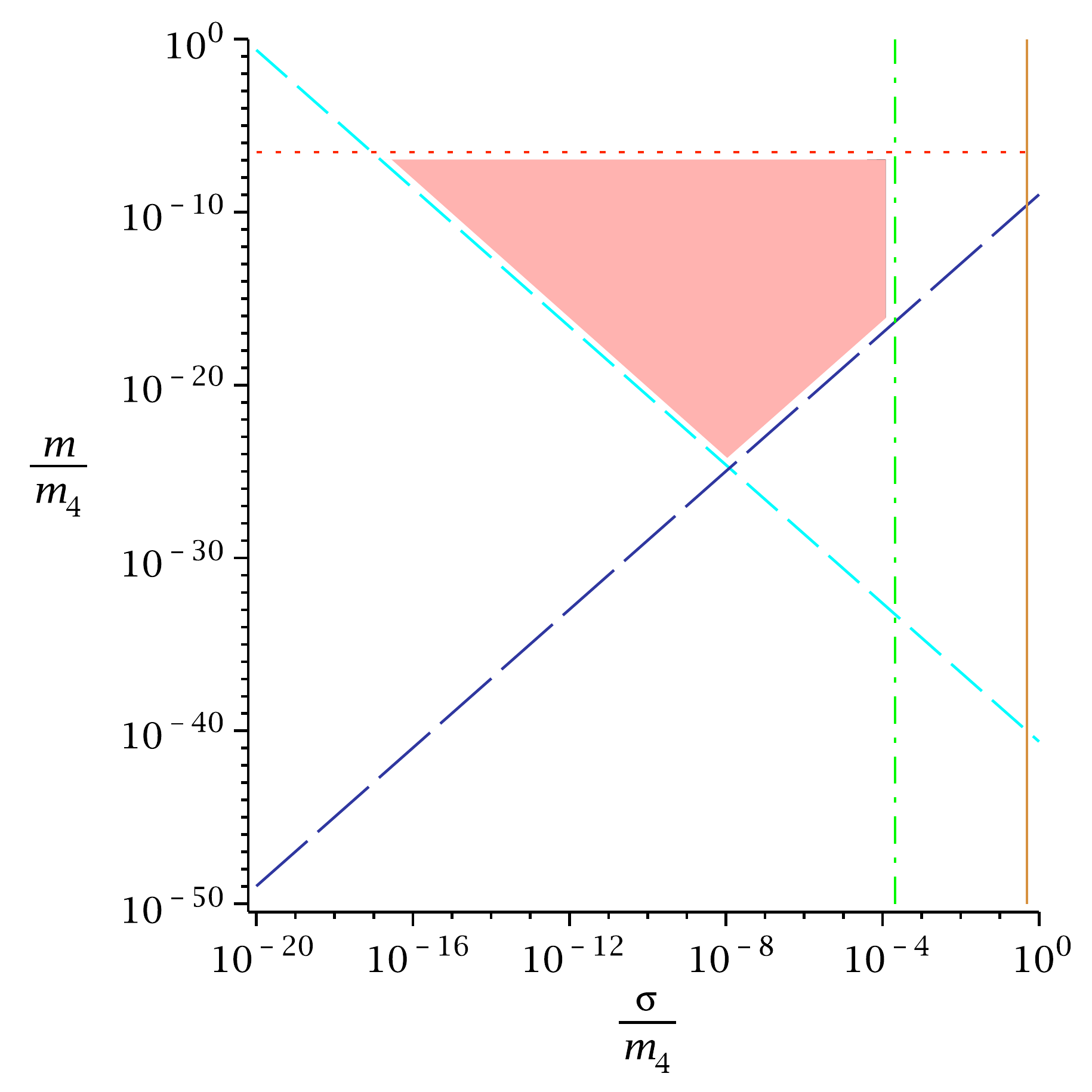}
\includegraphics[width=0.32\linewidth]{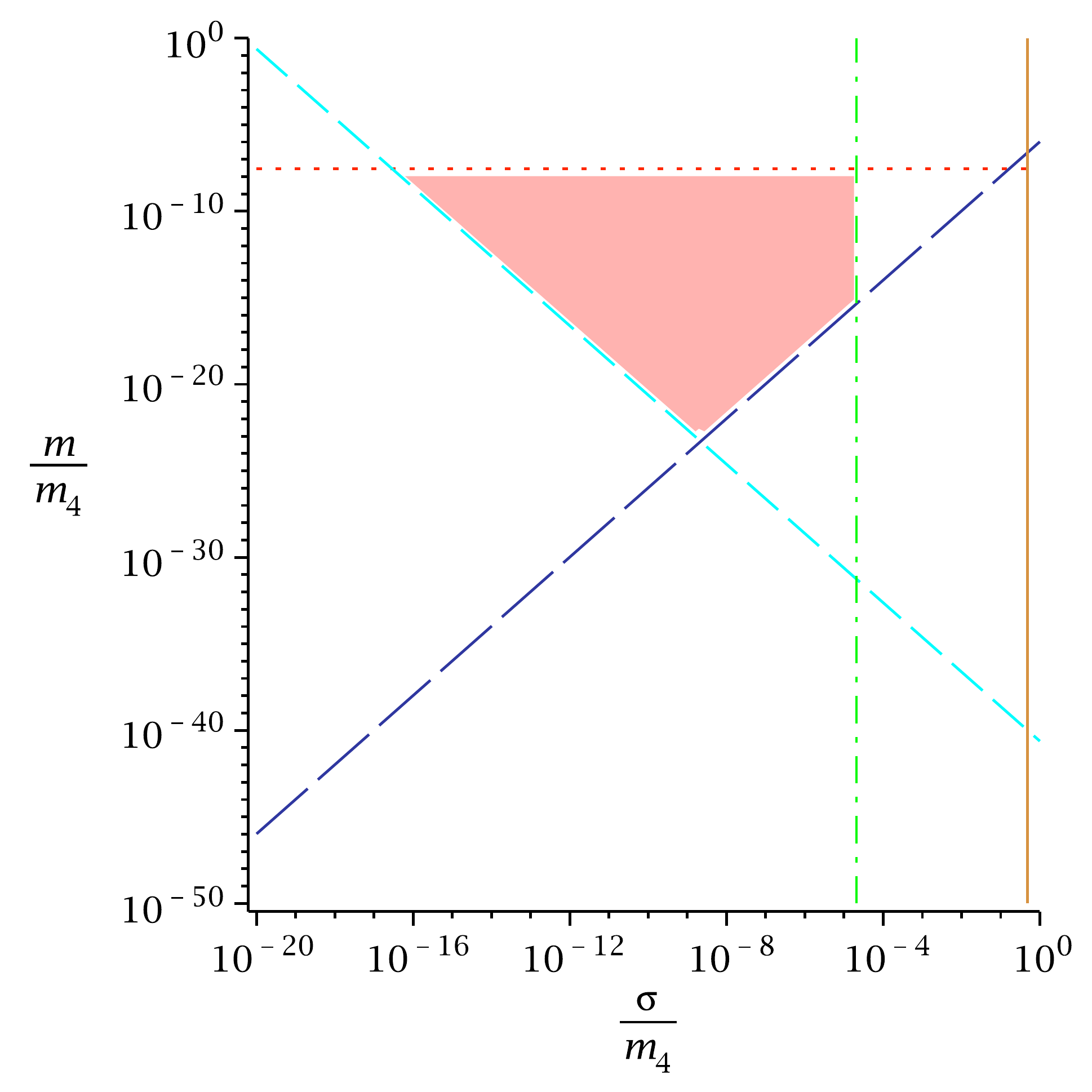}
\caption{Exponential potential with curvaton, showing constraints on the values of $\sigma$ and $m$ for different $\alpha$. From left to right, $\alpha=5$, $\alpha=50$ and $\alpha=500$. The viable space of parameters corresponds to the shaded area. The solid/orange line represents condition 1, the dotted/red line condition 2, the dashed/cyan line condition 3, the dash-dotted/green line condition 4 and the long-dashed/blue line condition 5. $N$ was set to 70. The change in $\alpha$ doesn't strongly affect the results.}
\label{exp}
\end{figure*}

Now, we want to evaluate these conditions for our two potentials. Conditions 1 and 3 are the same, as they don't depend on the shape of the potential. 

For exponential scalar field potentials, using expressions (\ref{ViVeexp}) and (\ref{Vendexp}), condition 2 can be worked out as follows:
\begin{equation}
\frac{m^{2}\sigma_{i}^{2}4\pi}{2\alpha^{2}\lambda} = \frac{\alpha^{2}m^{2}\sigma_{i}^{2}(N+1)^{2}}{6H_{(i)}^{2}m_{4}^{2}} \ll 1.
\end{equation}
So, we can write down the expression as
\begin{equation}
m^{2} \ll \frac{54\pi^{2}P_{\zeta}m_{4}^{2}}{\alpha^{2}(N+1)^{2}},
\end{equation}
where $P_{\zeta}$ is the observationally-known spectrum of the Bardeen potential, which has the value of $ P_{\zeta}= 2 \times 10^{-9}$ at the horizon scale.

In condition 4, we need to insert the value that $\lambda$ would take if there was no curvaton in the model. This can be taken from the COBE normalization and looks like \cite{sahni}:
\begin{equation}
\lambda_{\rm no \ \sigma}= \frac{2.3 \times 10^{-10} (8 \pi)^{3}m_{4}^{4}}{\alpha^{6} (N+1)^{4}}.
\end{equation}
$\lambda$ can be expressed as a combination of $\sigma_{i}$ and known parameters by using expressions (\ref{Hi}) and (\ref{ViVeexp})
\begin{equation}
H_{i}^{2}=H_{e}^{2}(N+1)^{2}=9\pi^{2}P_{\zeta}\sigma_{i}^{2}=\frac{\alpha^{4}\lambda(N+1)^{2}}{12\pi m_{4}^{2}} 
\label{lambdaexp1}
\end{equation}
\begin{equation}
 \lambda=\frac{108 \pi^{3} P_{\zeta}\sigma_{i}^{2}m_{4}^{2}}{\alpha^{4}(N+1)^{2}}.
 \label{lambdaexp}
\end{equation}
Condition 4 is then just
\begin{equation}
\sigma_{i}^{2} \ll \frac{1 \times 10^{-9}m_{4}^{2}}{P_{\zeta}\alpha^{2}(N+1)^{2}}.
\end{equation}

Condition 5, when expressions (\ref{Hi}), (\ref{ViVeexp}), (\ref{Vendexp}), (\ref{rhoGW}) and (\ref{hGW1}) are plugged in, looks like:
\begin{equation}
\frac{4}{\pi m_{4}^{2}}\alpha^{2}(N+1)9P_{\zeta}\sigma_{i}^{2}\left(\frac{a_{\rm (eq)}}{a_{(k)}}\right)^{2} \ll 1
\label{condition51}
\end{equation}
which, using Eqs.~(\ref{Vendexp}) and (\ref{equivalence}), can be written as (here we use the assumption that $H_{(k)}=H_{(e)}$)
\begin{equation}
\frac{4}{\pi m_{4}^{2}}\alpha^{2}(N+1)9P_{\zeta}\sigma_{i}^{2} \left(\frac{3 m_{4}}{4\pi m \sigma_{i}^{2}}\right)^{2/3} \left(\frac{\alpha^{2} \sqrt{\lambda}}{\sqrt{12\pi}}\right)^{2/3}\ll 1.
\label{condition52}
\end{equation}
Substituting in the expression for $\lambda$, Eq.~(\ref{lambdaexp}), and rearranging, the condition is simply
\begin{equation}
m \gg \frac{5 \times 10^{2}}{\pi^{3/2} m_{4}} \alpha^{3}\sqrt{(N+1)} P_{\zeta}^{2} \sigma_{i}^{2}.
\label{condition53}
\end{equation}

\begin{figure}[t]
\includegraphics[width=0.9\linewidth]{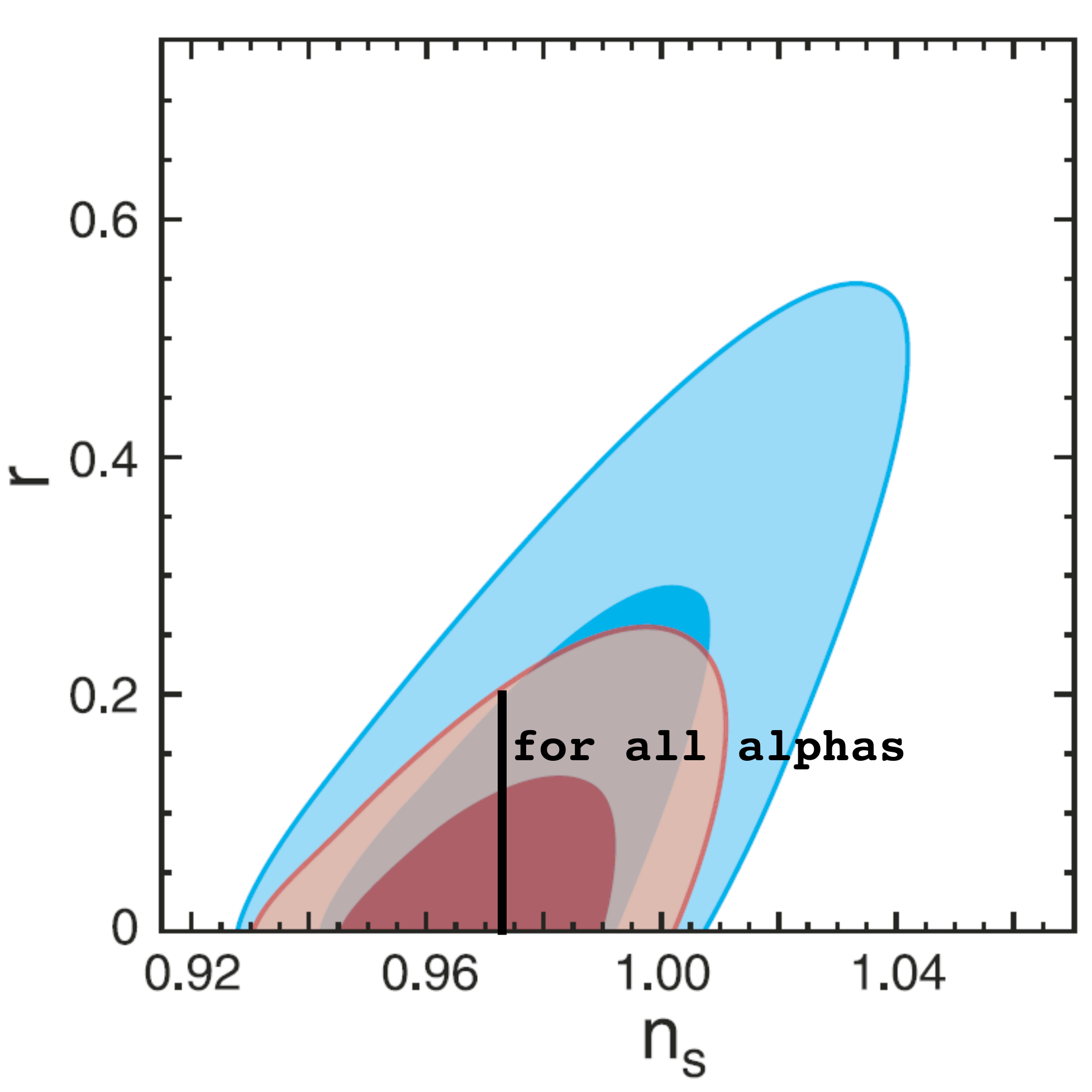}
\caption{Exponential potential with curvaton. Comparison of the values of $n_{\rm s}$ and $r$ with observations from the WMAP five-year data. The data shown is as in Fig.~\ref{wmapinvnoc}. The parameter $r$ can take different values depending on $\sigma$. The interval of possible values of $r$ is the same for all values of $\alpha$. We can see that this model fits data successfully. Adapted from Ref.~\cite{Komatsu5yrWMAP}.}
\label{wmapexp}
\end{figure}

\begin{figure*}[t]
\includegraphics[width=0.32\linewidth]{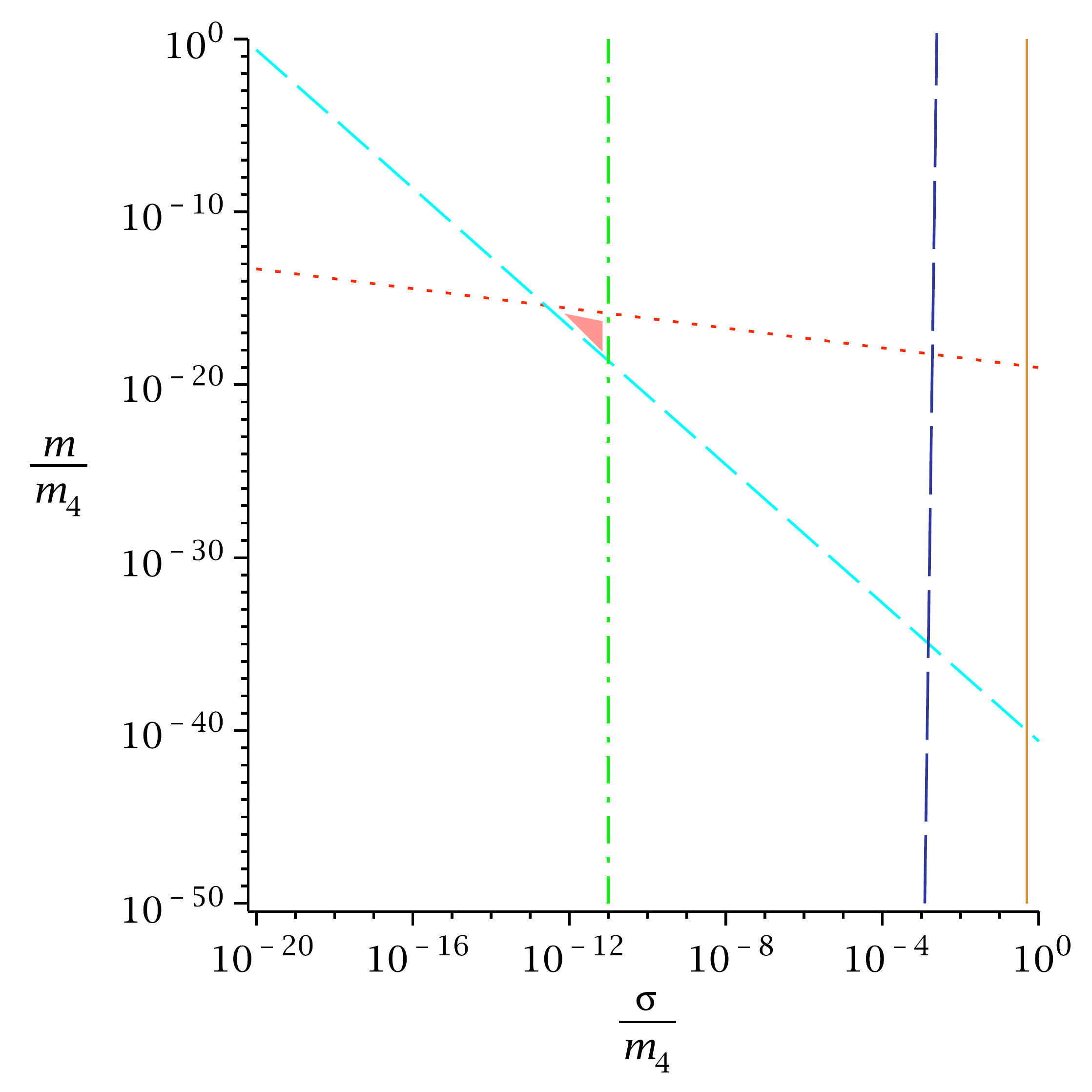}
\includegraphics[width=0.32\linewidth]{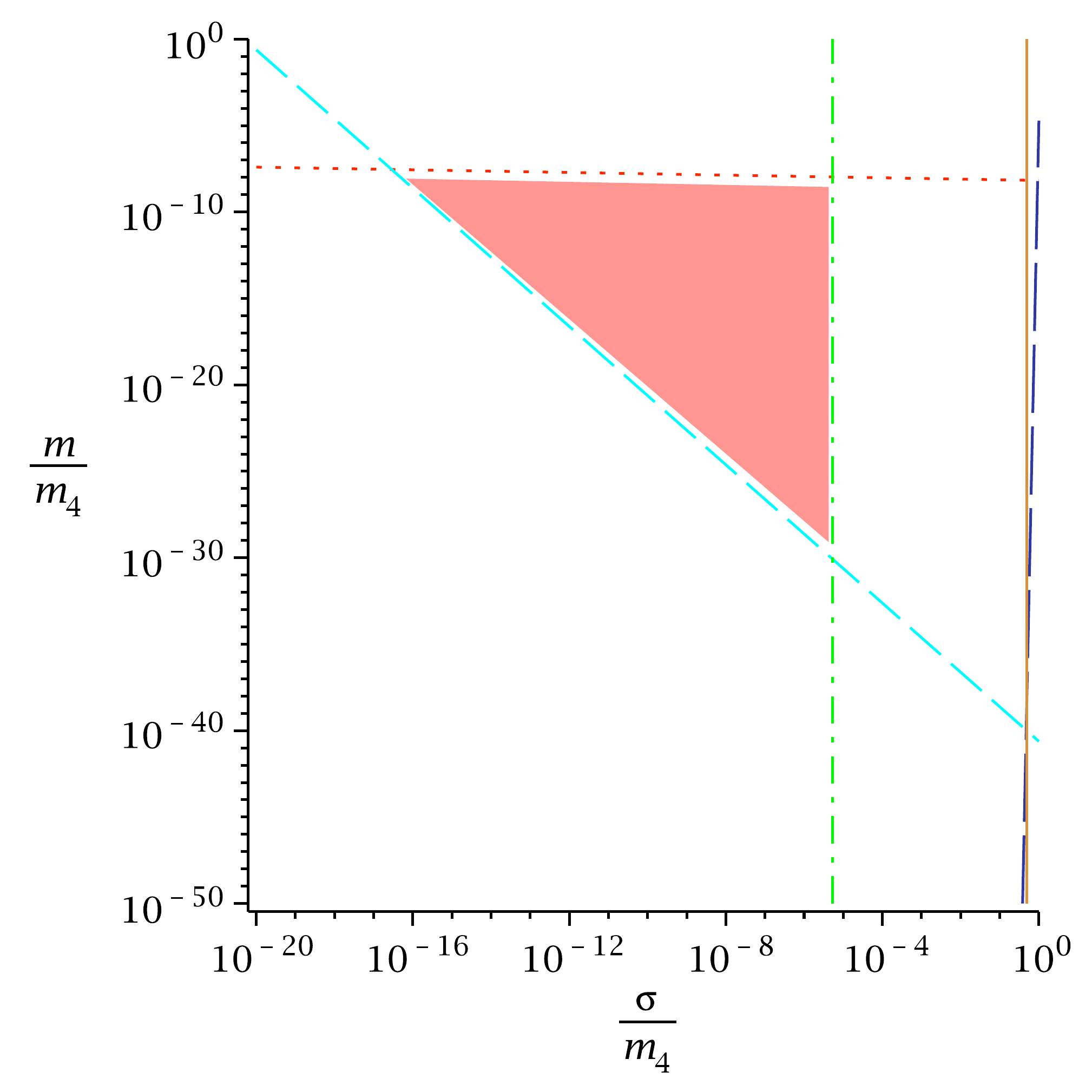}
\includegraphics[width=0.32\linewidth]{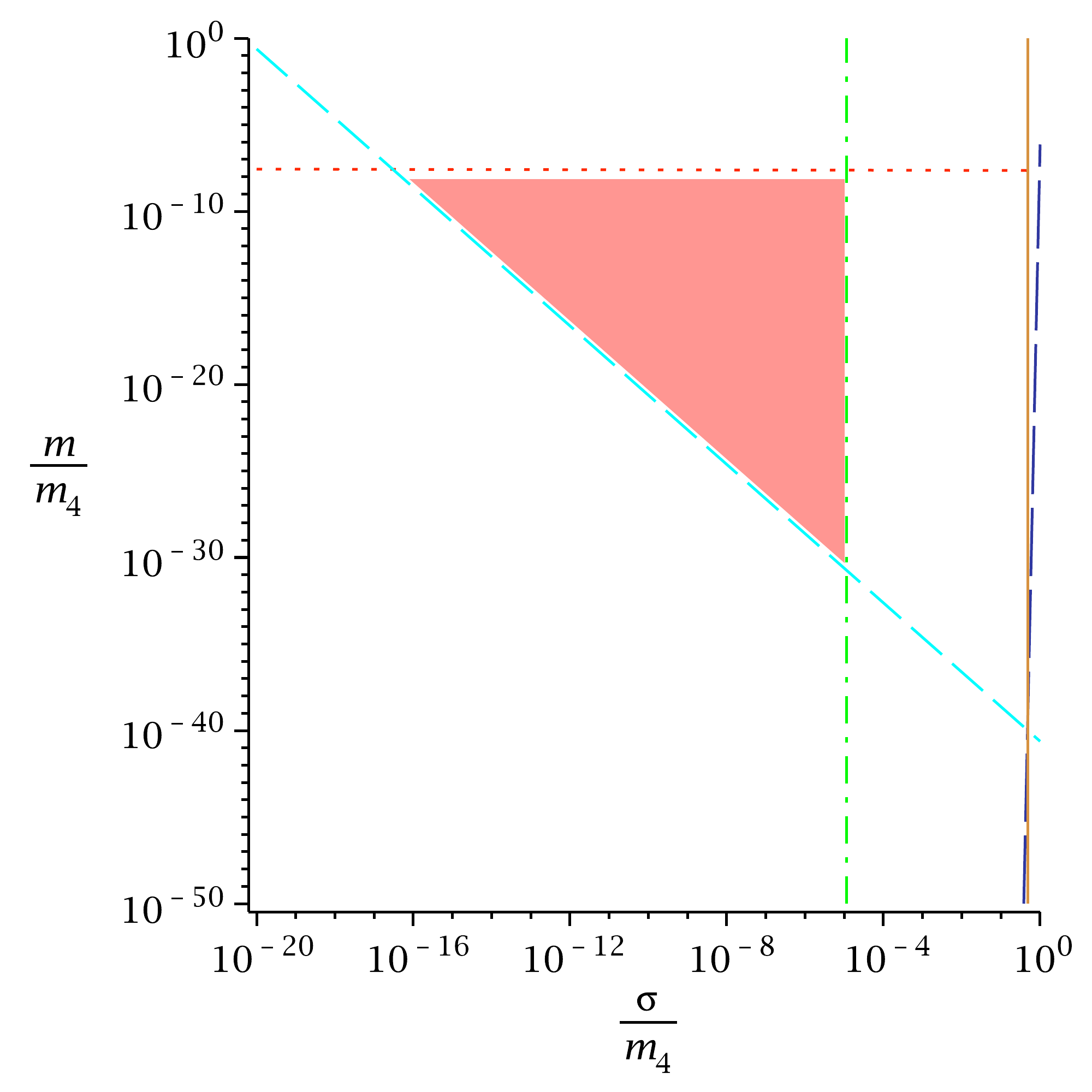}
\caption{Inverse power-law potential with curvaton, showing constraints on the values of $\sigma$ and $m$ for different $\alpha$. From left to right, $\alpha=5$, $\alpha=50$ and $\alpha=500$. The viable space of parameters corresponds to the shaded area. The solid/orange line represents condition 1, the dotted/red line condition 2, the dashed/cyan line condition 3, the dash-dotted/green line condition 4 and the long-dashed/blue line condition 5. $N$ was set to 70 and $V_{0}$ to  $10^{-100} m_{4}^{4}$. We can see how the change in $\alpha$, keeping $V_{0}$ fixed, strongly affects the results.}
\label{inv}
\end{figure*}

Figure \ref{exp} shows these constraints in the $m$--$\sigma$ plane. There is a dependence on $\alpha$, but as can be seen it is very weak. There is a comfortable parameter space where viable inflation can occur. In Fig.~\ref{wmapexp} we plotted the result for $n_{\rm s}$ and $r$ for this case. The parameter $n_{\rm s}$ is independent of $\alpha$ and $\sigma$. The parameter $r$ depends on $\sigma$ but the interval of possible values it can take is the same for each value of $\alpha$. This model of inflation is in very good agreement with observations. In the next section we will see whether it can also produce a satisfactory quintessence. \\

For inverse power-law potentials, we use always the approximate expression for $V_{(e)}$, Eq.~(\ref{Vendinv}).  Using this function, and following the steps as for Eq.~(\ref{lambdaexp1}),  $\lambda$ can be expressed as
\begin{equation}
\lambda=\left(\frac{3 \times 10^{4} \pi \, P_{\zeta} m_{4}^{2}\sigma_{i}^{2}V_{0}^{4/(\alpha-2)}}{16\alpha^{4}(N+1)^{2\alpha/(\alpha-2)}}\right)^{\frac{\alpha-2}{\alpha+2}}.
\label{lambdainv}
\end{equation}
So, condition 2 is
\begin{equation}
m^{2} \ll 0.12\left(\frac{2 \times 10^{3} \pi m_{4}^{2} P_{\zeta}}{(N+1)^{2\alpha/(\alpha-2)}}\right)^{\frac{\alpha}{\alpha+2}}\alpha^{\frac{-2(\alpha-2)}{\alpha+2}}V_{0}^{\frac{2}{\alpha+2}}\sigma_{i}^{\frac{-4}{\alpha+2}}.
\end{equation}

For condition 4 we need to know what the value of $\lambda$ would be without the presence of the curvaton. It looks like \cite{sahni}
\begin{equation}
\lambda_{{\rm no} \ \sigma}= V_{0}^{\frac{6}{\alpha+4}}m_{4}^{\frac{4\alpha-8}{\alpha+4}}\left(\frac{P_{\zeta}}{(N+1)^{4}}\right)^{\frac{\alpha-2}{\alpha+4}}\left(\frac{4\pi}{\alpha^{2}}\right)^{\frac{3\alpha}{\alpha+4}}.
\end{equation}
Using expressions  (\ref{Vendinv}) and (\ref{lambdainv}), the condition can be written as
\begin{align}
&\sigma_{i} \ll 6 \times 10^{-3} \Big(4^{5\alpha^{2}+10\alpha -16} V_{0}^{2\alpha-4} m_{4}^{2\alpha(\alpha-2)} P_{\zeta}^{-2\alpha+4} \nonumber\\
&\quad (N+1)^{-2\alpha^{2}+8\alpha+16} \pi^{2\alpha^{2}+4\alpha+8} \alpha^{-2\alpha^{2}-4\alpha-32}\Big)^{\frac{1}{2(\alpha-2)(\alpha+4)}}.
\end{align}

\begin{figure}[t]
\includegraphics[width=0.9\linewidth]{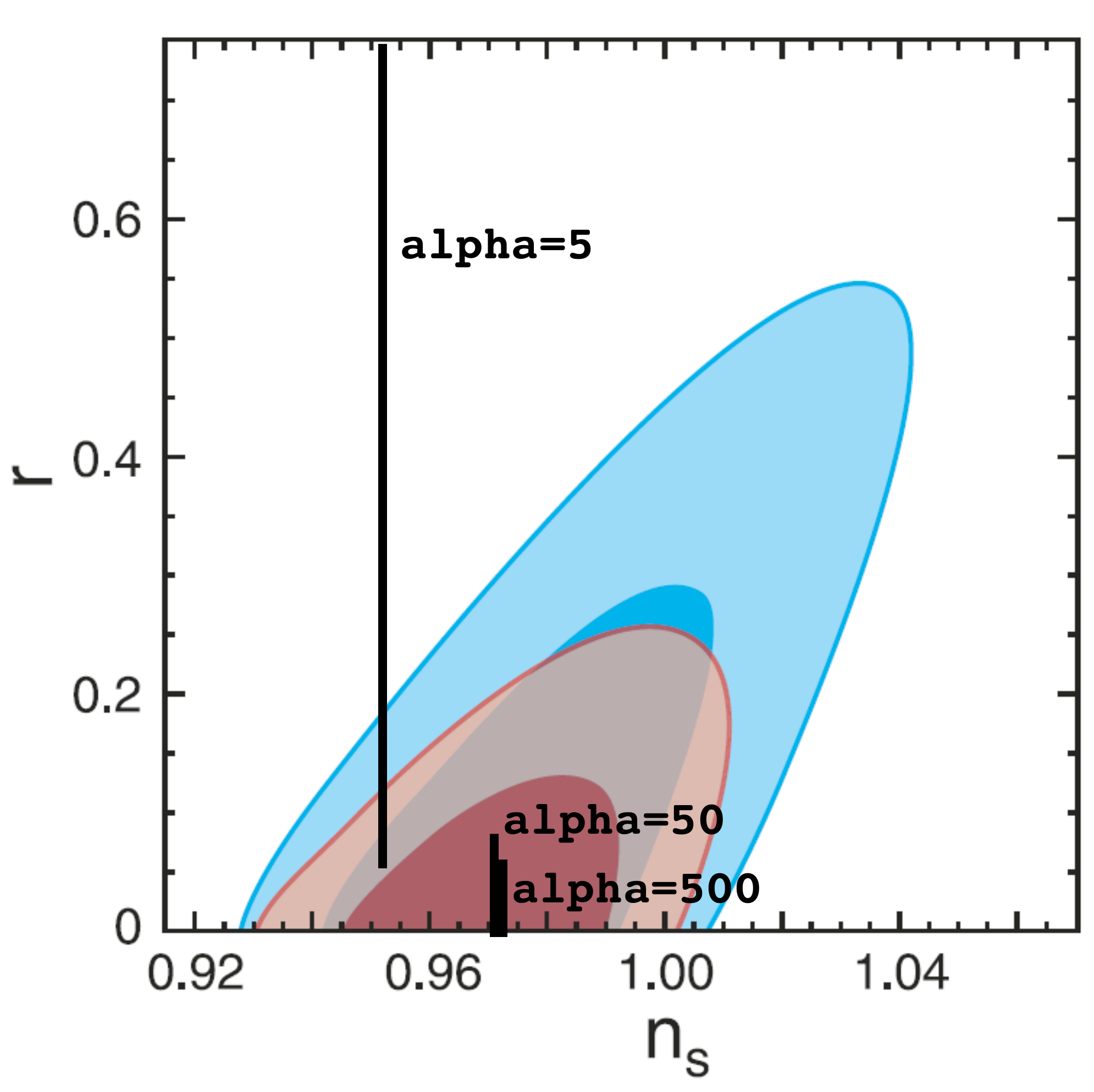}
\caption{Inverse power-law potential with curvaton.  Comparison of the values of $n_{\rm s}$ and $r$ with observations from the WMAP five-year data, for different $\alpha$. The data shown is as in Fig.~\ref{wmapinvnoc}. For small $\alpha$ we can see that the parameter $r$ can be too high compared to the observations. We can then impose an extra constraint on the combination of $\sigma$, $V_{0}$ and $\alpha$. We can see that this model presents a big space of parameters that fits data successfully. Adapted from Ref.~\cite{Komatsu5yrWMAP}.}
\label{wmapinv}
\end{figure}

At this stage, $V_{0}$ is a free parameter, which will be fixed when we impose that we want quintessence to start domination today. However, $V_{0}$ is not completely free, since the viability of the model depends strongly on the combination of this parameter and $\alpha$. This can be seen by looking at our condition. Increasing $V_{0}$ (with all other variables fixed) increases the right-hand side of the inequality. Schematically, this corresponds to 
the dash-dot line of Fig.~\ref{inv} moving towards higher values of $\sigma$, increasing the space of parameters available. On the other hand, if both $V_{0}$ and $\alpha$ were really small, the model would be unable to create good inflation. In Figs.~\ref{inv} and \ref{wmapinv}, $V_{0}$ is set to $10^{-100} m_{4}^{4}$ because it is a good initial guess for acceleration to happen today.  This issue will be analyzed in more detail next section when we impose the correct value on $V_{0}$ based on quintessence arguments.

To get condition 5, we just follow the exact same steps as in Eqs.~(\ref{condition51}), (\ref{condition52}) and (\ref{condition53}), leading to
\begin{align}
&m \gg \Big(1.2^{3(\alpha^{2}-4)}(2 \times 10^{3})^{(\alpha+8)(\alpha-2)}\pi^{6(\alpha-2)}\alpha^{6(\alpha-2)^{2}}\nonumber\\
&\quad V_{0}^{-6\alpha+12}(N+1)^{\alpha^{2}-10\alpha}m_{4}^{(-2\alpha+8)(\alpha-2)}\sigma_{i}^{(4\alpha+20)(\alpha-2)}\nonumber\\
&\quad P_{\zeta}^{(4\alpha+14)(\alpha-2)}\Big)^{\frac{1}{2(\alpha^{2}-4)}}.
\end{align}

Figure \ref{inv} shows these five conditions for the inverse power-law potential in the $m$--$\sigma$ space. In Fig.~\ref{wmapinv} we compare the range of values that $n_{\rm s}$ and $r$ can take with observations. We can see that the result is in good agreement with observations, although, for very small $\alpha$, some values of $\sigma$ will give rise to too large an $r$. We can then impose a new constraint on the combination of the values of $\sigma$, $V_{0}$ and $\alpha$ based on observations: $r$ needs to be smaller than approximately $0.2$ \cite{Komatsu5yrWMAP}.

\subsection{Quintessence}

In this section we will constrain the parameter space based on the observational data we have for quintessence. This is simply that today the dark energy density is approximately $\Omega_{\phi} \approx 0.75$ and its equation of state has $w < -0.8$ \cite{Komatsu5yrWMAP, Komatsu7yrWMAP}. A successful model of quintessential inflation is then one for which the conditions at the end of inflation allow the scalar field to have these characteristics today. In other words, we want the scalar field to leave inflation at such a point in the potential and with such a velocity, that it will enter in a regime, being it tracking or thawing, that will resolve in a satisfactory domination today. 

Generally, smaller field values and larger velocities will increase the duration of kination and can force the scalar field to enter a frozen regime that can only redominate through a thawing process. On the other hand, for the case of no velocity, the field always finds its tracking solution, it being an attractor. For each $\alpha$, the free variables that we can tune in order to change the conditions at the end of inflation are $\lambda$ and $V_{0}$.  These completely determine $\phi_{e}$ and $\dot{\phi}_{e}$ as we can see by setting the slow-roll parameter $\epsilon$ to 1: in the standard formalism, $\epsilon=1$ determines $\phi_{e}$ \cite{maartens} and, in the Hamilton--Jacobi formalism, $\epsilon_{H}=1$ determines $\dot{\phi}_{e}$ \cite{LPB}. For example, for an inverse power-law potential
\begin{equation}
\epsilon = \frac{m_{4}^{2} V'^{2} \lambda}{4\pi V^{3}} \Rightarrow \phi_{e}=\left(\frac{\alpha^2 \lambda}{4\pi V_{0}}\right)^{\frac{1}{2-\alpha}}
\label{phiend}
\end{equation}
\begin{equation}
\epsilon_{H}= \frac{3\dot{\phi}^{2}/2}{\dot{\phi}^{2}/2 + V} \Rightarrow \dot{\phi}_{e}= \lambda^{\frac{\alpha}{2\alpha-4}}V_{0}^{-\frac{1}{\alpha-2}} 0.24 \alpha .
\label{phidotend}
\end{equation}

By choosing $\phi_{e}$ and $\dot{\phi}$, for each $\alpha$, we can calculate the $w$ of the scalar field today. We did this by performing a numerical integration of the equations of motion of the system from the end of inflation until today. In our calculation we fixed $\Omega_{\phi}=0.75$ today. Given a certain set of initial conditions, the program had the freedom to rescale $V_{0}$ as many times as necessary to ensure that today indeed $\Omega_{\phi}=0.75$. Once this condition was satisfied, one of the outputs of the calculation was $w$.  

The integration was done with respect to the number of $e$-foldings, so instead of Eq.~(\ref{phidotend}) we use
\begin{equation}
\frac{d\phi}{dN}=\frac{d\phi}{dt} \frac{dt}{dN}=\frac{2}{\alpha} \left( \frac{V_{0}}{\lambda} \right)^{\frac{1}{\alpha-2}}.
\label{dphidN}
\end{equation}
Conveniently both $\phi_{e}$ and $d\phi/dN$ depend only on the ratio $\lambda/V_{0}$ and on $\alpha$.\\

\begin{figure}[t]
\includegraphics[width=0.95\linewidth]{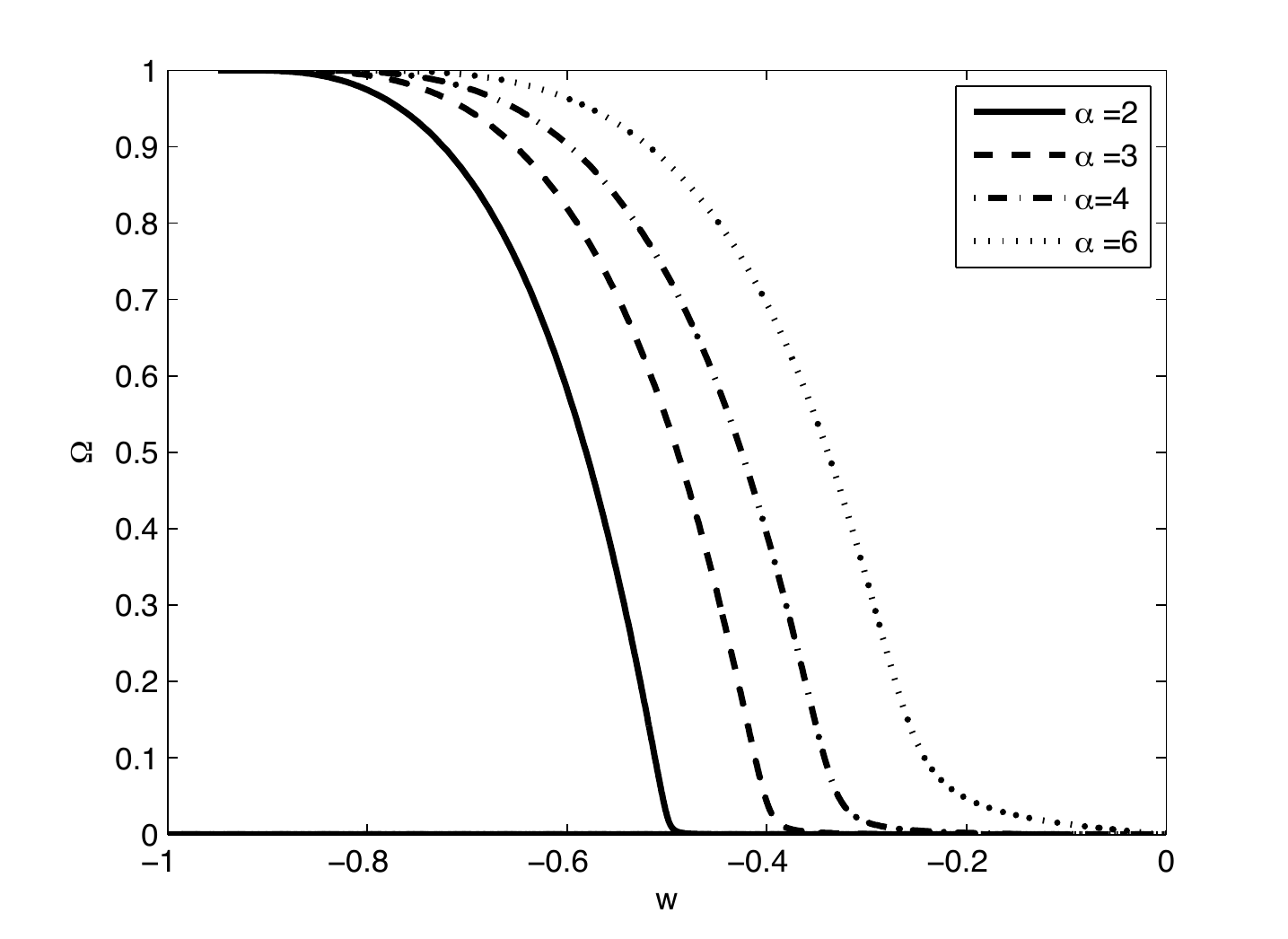}
\caption{$\Omega$ and $w$ for different $\alpha$'s for the tracking solution of an inverse power law. We can see that for $\alpha > 2$, as it is required for the system to inflate, it is impossible to match the observational constraints.}
\label{tracking}
\end{figure}

To check if any tracking solution would give a good quintessence behaviour, and since we already know that for the exponential case it doesn't, we run the calculation for inverse power-law potentials with $d\phi/dN$ at end of inflation equal to zero. For any $\phi$, because the solution is an attractor, the scalar field found the tracking regime. The conclusion is that no tracking solution, for any $\alpha$, could create a quintessence domination with the characteristic that we observe. The results are presented in Fig.~\ref{tracking} and match those of Ref.~\cite{huey}. In particular, for very big $\alpha$'s, the inverse power-law potential approaches the exponential one, and the scalar field never manages to dominate. Very small $\alpha$ like 1, which could give a viable quintessence, aren't interesting for us since we know that the system doesn't inflate for such a potential.

To understand the true behaviour of the system, we run the calculation with the initial conditions as in the expressions (\ref{phiend}) and (\ref{dphidN}). By choosing the ratio $\lambda/V_{0}$, since the program will give the value of $V_{0}$, we are fixing $\lambda$. We run it for inverse power laws, and we understand what occurs for exponential potentials by analyzing the large $\alpha$ limit. 

Firstly, we concluded that for $\lambda/V_{0} < 1$ the model doesn't work because the scalar field never stops dominating.

We also see that for small $\alpha$'s and small $\lambda/V_{0}$ the scalar field freezes and dominates at late times through a thawing process. This only happens for $\alpha<12$. However, only for $\alpha \leq 6$ we can actually achieve an equation of state $w< -0.8$ today. For greater $\alpha$'s, $w$ is too big.

\begin{figure}[t]
\includegraphics[width=0.85\linewidth]{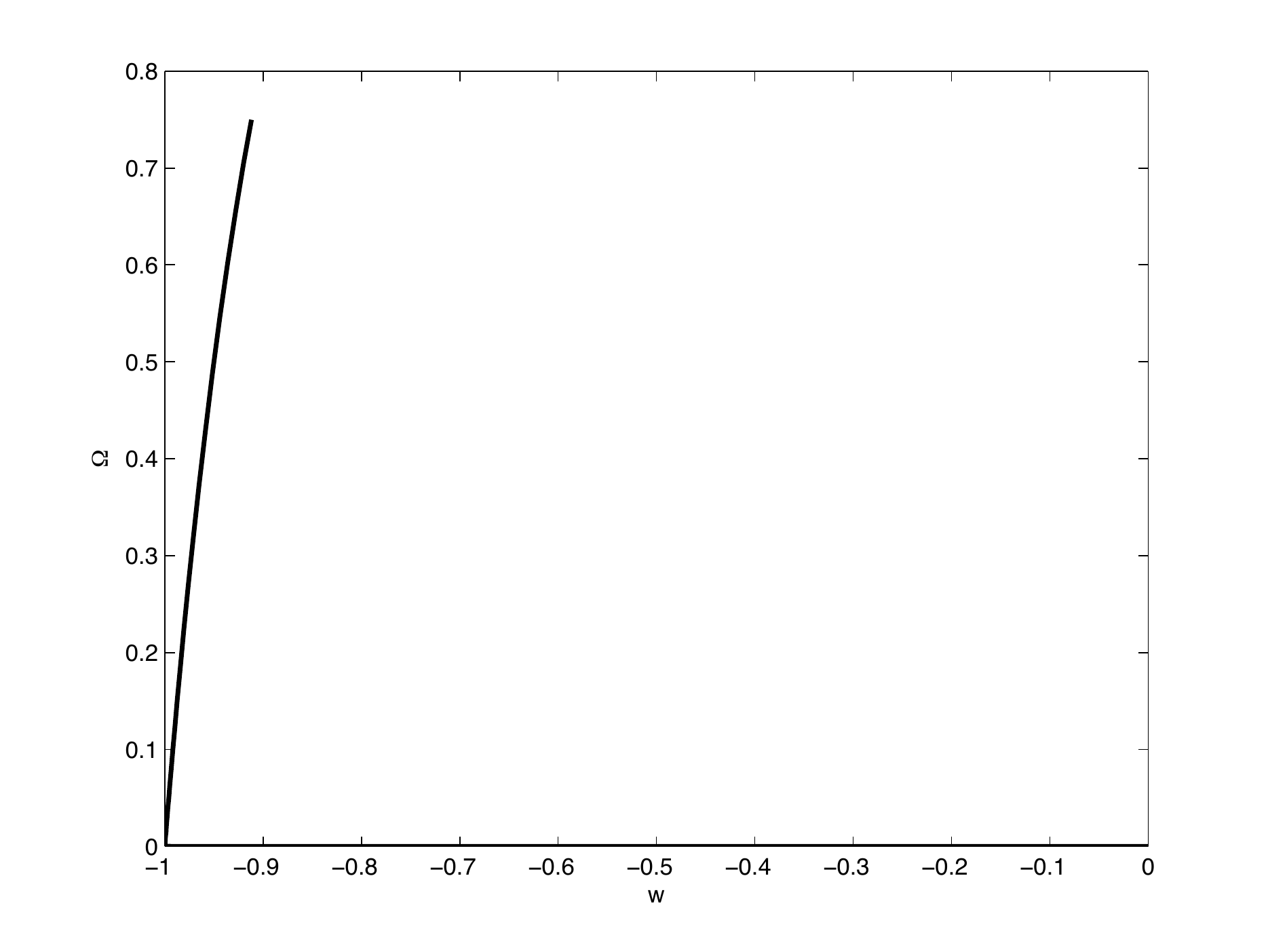}\\
\includegraphics[width=0.85\linewidth]{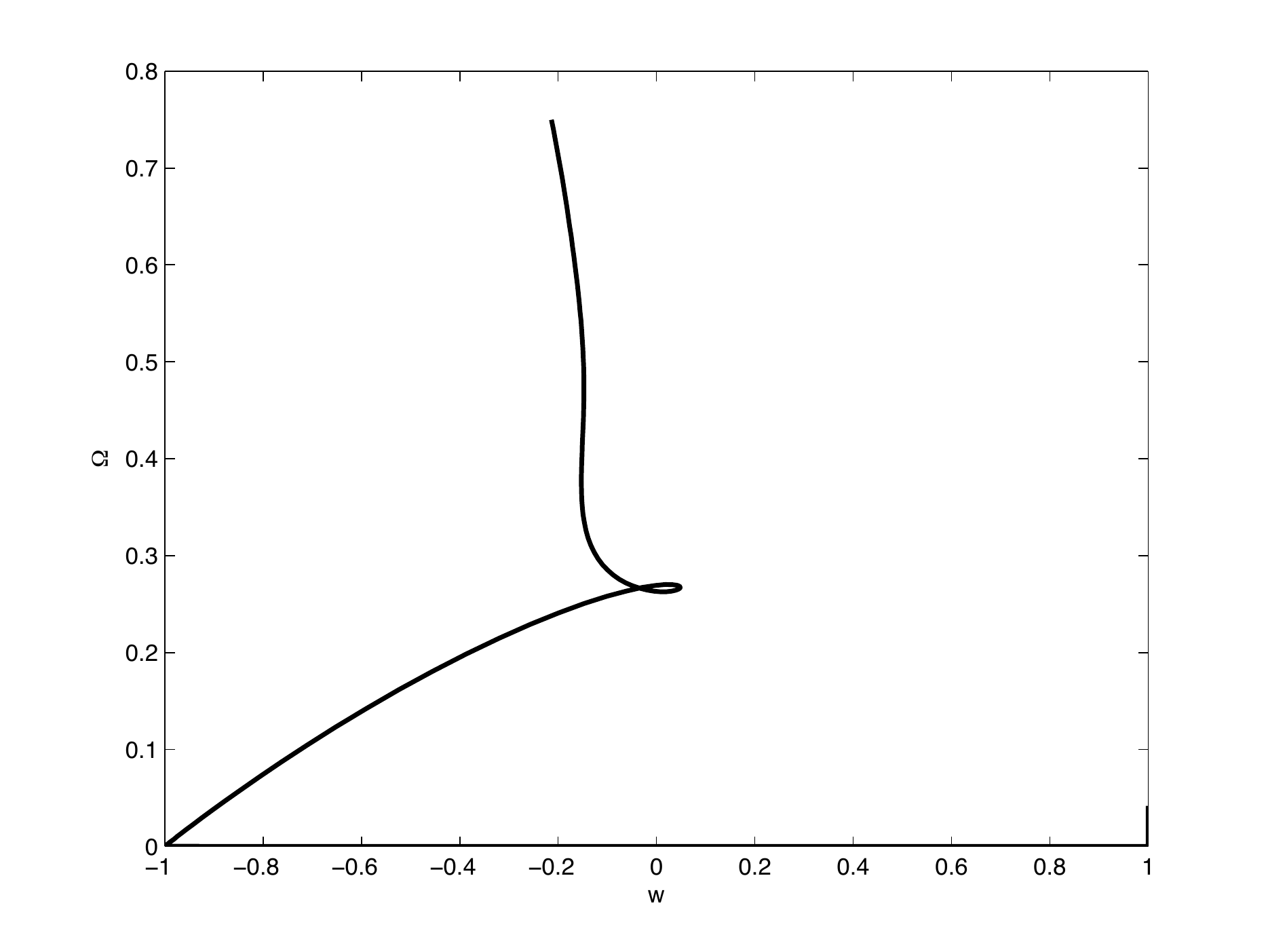}
\caption{Examples of different evolutions of the scalar field. In the upper plot, $\alpha=5$. We see that the field is initially frozen and starts evolving until domination. This model is in good agreement with observations. In the lower plot, $\alpha=20$. We see that the field again begins in a frozen regime, but during its evolution, it finds the tracking solution. In this case, there is a disagreement with observations. In both plots, $\lambda/V_{0}=1$.}
\label{thawVstrack}
\end{figure}

As $\alpha$ increases it becomes harder for the field to have enough speed to freeze until domination by thawing, and it finds the tracking solution before dominating. Figure \ref{thawVstrack} shows two examples for the field thawing and finding its tracking solution before domination, respectively. 

For very large $\alpha$, as expected, the field never manages to dominate at late times. We already knew that once the field finds its tracking solution, the model fails.

We conclude, then, that for our conditions at the end of inflation, only for very particular cases of very small $\alpha$ and $\lambda/V_{0}$ are the observational constraints of quintessence satisfied. Now we need to check if these constraints are in agreement with the ones from inflation.

\section{Quintessential inflation}

A satisfactory quintessence, as a manifestation of the inflaton at late times, is only achieved, as mentioned, for inverse power law potentials with $2 \leq \alpha \leq 6$ with low values of $\lambda/V_{0}$ (though bigger than 1). But for the full model to work, these constraints must be in agreement with inflation itself. 

When we fix $\lambda/V_{0}$ in the calculation, since $V_{0}$ is an output value, we are fixing $\lambda$. There are two tests we can do to check if inflation works successfully in the cases where quintessence does. First we can substitute the output $V_{0}$ in the conditions for inflation and create a plot as in Figs.~\ref{exp} and \ref{inv} to see if there is still a region of suitable parameters. If there is, the second test we can make is to substitute the value of $\lambda$ in expression (\ref{lambdainv}) and see if the value of $\sigma_{i}$ lays in such a region. 

\begin{figure}[t]
\includegraphics[width=0.95\linewidth]{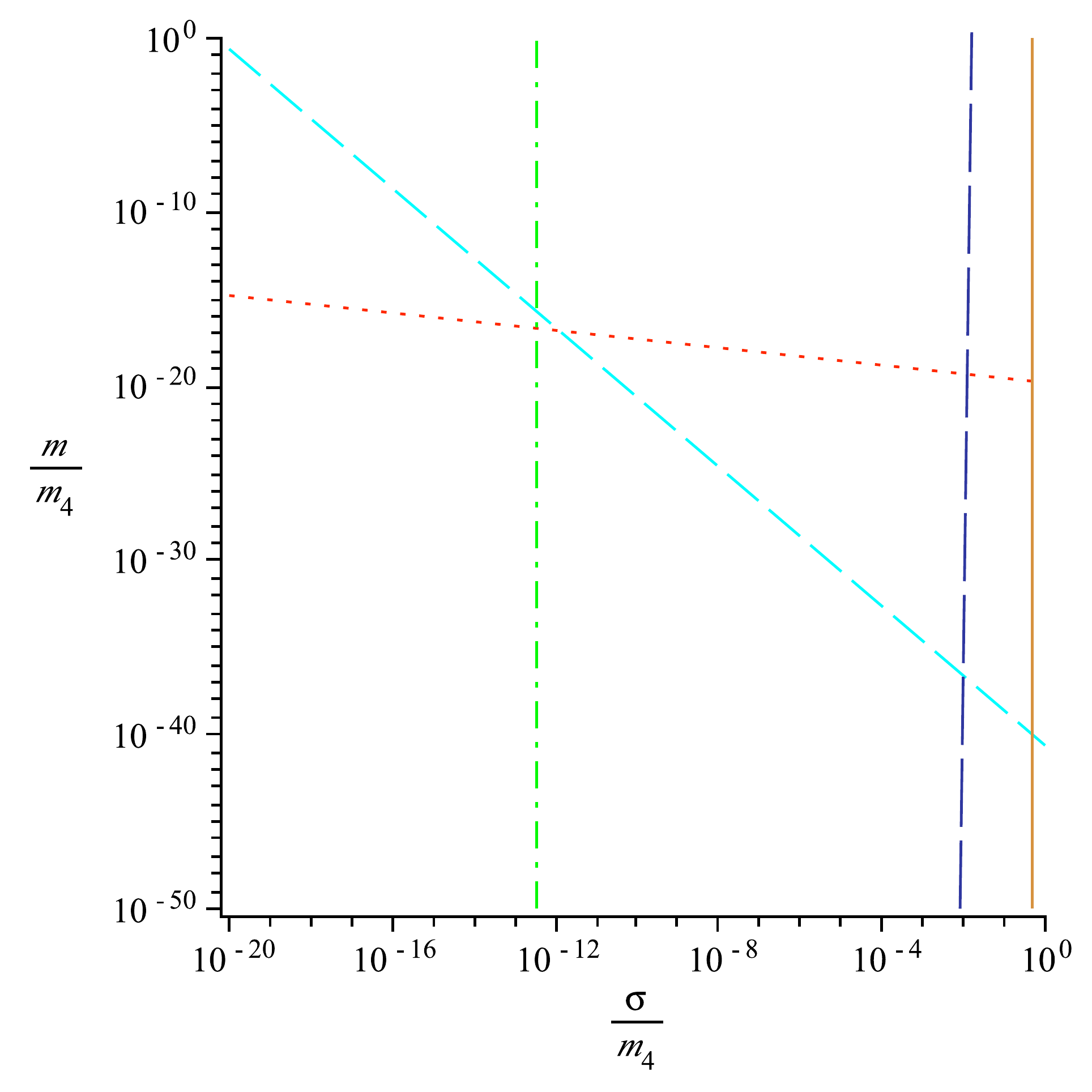}
\caption{Parameter space for $\alpha=6$ and $V_{0}=10^{-120}m_{p}^{4}$. We see that there is no possibility of successful inflation (the allowed parameter space is to the left of the dash-dotted/green line, above the dashed/cyan line and below the dotted/red line). For smaller values of $\alpha$ the discrepancy increases. }
\label{fail}
\end{figure}

In our case, $V_{0}$ is of the order of $10^{-120} m_{4}^{4}$ for all the possible $\alpha$'s. This is a very low value. It implies that inflation would end at an energy scale of $10^{-120} m_{4}^{4}$, which is much lower than the energy scale for nucleosynthesis. This incompatibility is clear in our plot of the parameter space. Figure \ref{fail} shows how by implementing the real value of $V_{0}$ for $\alpha=6$, the parameter space disappears. This fact is even more drastic for smaller values of $\alpha$.

Hence we conclude that we cannot get quintessential inflation for the potentials studied. This is our main conclusion.

Now, we can discuss how this result would change for other potentials. In this paper, we chose to study the simplest potentials that have interesting late-time behaviours. More complex potentials with more freedom of parameters, when picked correctly, can be set to fit observations. In particular, if the potential is composed of two different regimes at early and late times, as in Ref.~\cite{PV}, it is always possible to fit the observations, just by choosing good potentials and by setting exactly where the late-time behaviour should start. Since our objective was to try to motivate the value of $\Omega_{\phi} \approx 0.75$ today by the conditions of the Universe at the end of inflation, such models aren't in the scope of our analysis. 

To end our discussion, although quintessential inflation is impossible to achieve in our framework, we want to highlight the fact that it is still possible to get inflation. In such a case, for both potentials, the system would have to have a curvaton to  give rise to the density perturbations and satisfy the constraints shown in Figs.~\ref{exp} and \ref{inv}. Additionally, to prevent it coming to redominate too early, the inflaton would have to decay completely through some sort of reheating process, for instance a hybrid inflation mechanism, leading to considerable further modelling complications. In this case, another explanation of the dark energy would need to be sought.

\section{Conclusions}

We have presented a comprehensive series of models seeking to uncover the possibility of combining early Universe inflation and present-day quintessence into a single field with a simple potential. Such a model would give obvious benefits in simplifying the structure of the cosmological model, but must face much stronger observational challenges than models where separate components are responsible for each accelerated period.

The range of models we have discussed is quite extensive. Having adopted the braneworld in order to have a natural end to inflation which does not require a feature in the potential, we then considered two substantially different reheating mechanisms --- gravitational particle production and curvaton reheating --- as well as two different regimes of quintessence, tracking and thawing. In practice, gravitational particle production fails both to match the primordial spectra measured by WMAP and to give success nucleosynthesis due to excessive short-scale gravitational waves, forcing us to the curvaton case. Likewise, the conditions that inflation enforces on the potential immediately rule out the tracking models, as either the field fails to dominate at all at late times, or does so with an equation of state too far from $-1$. We are thus forced to the thawing regime, as first pointed out in Ref.~\cite{huey}. It is then possible to find models capable of satisfying either the inflationary part of the picture, or the quintessence part. Sadly, however, we have found no models capable of satisfying both, the combined observational constraints being failed by a large margin. 

Given the comprehensiveness of the models we have considered, we feel that current observations leave no room for quintessential inflation from simple potentials, with models assembled from the various known mechanisms of inflation, reheating, and quintessence. Quintessential inflation does, of course, remain possible provided one manipulates the potential into just the required shape to bring the different regimes together, as in the original proposal of Peebles and Vilenkin \cite{PV}. To do so does not however appear any more attractive than maintaining two separate sectors, nor does it seem amenable to future observational tests.


\begin{acknowledgments}
M.D.\ was supported by FCT (Portugal) and A.R.L.\ by STFC (UK). 
\end{acknowledgments}



\begin{thebibliography}{}

\bibitem{quevedo} F. Quevedo,
Class. Quant. Grav. {\bf19}, 5721 (2002),
 \texttt{hep-th/0210292}. 

\bibitem{PV} P. J. E. Peebles and A. Vilenkin, Phys. Rev. D{\bf 59}, 063505 (1999), 
\texttt{astro-ph/9810509}.

\bibitem{RS2} L. Randall and R. Sundrum, Phys. Rev. Lett. {\bf 83}, 4690 (1999), 
\texttt{hep-th/9906064}.

\bibitem{langlois}P. Binetruy, C. Deffayet, and D. Langlois,
Nucl. Phys. B{\bf 565}, 269 (2000),
\texttt{hep-th/9905012};
P. Binetruy {\it et al.},
Phys. Lett. B{\bf 477}, 285 (2000),
\texttt{hep-th/9910219}.

\bibitem{maartens}R. Maartens {\it et al.},
Phys. Rev. D{\bf 62}, 041301 (2000),
\texttt{hep-th/9912464}.

\bibitem{steep}E. J. Copeland, A. R. Liddle, and J. E. Lidsey,
Phys. Rev. D{\bf 64}, 023509 (2001),
\texttt{astro-ph/0006421}.

\bibitem{gravprod} L. H. Ford, Phys, Rev. D{\bf 35}, 2955 (1987); B. Spokoiny, Phys. Lett. B{\bf 315}, 40 (1993).

\bibitem{huey}G. Huey and J. E. Lidsey,
Phys. Lett. B{\bf 514}, 217 (2001),
\texttt{astro-ph/0104006}.

\bibitem{sahni}V. Sahni, M. Sami, and T. Souradeep,
Phys. Rev. D{\bf 65}, 023518 (2002),
\texttt{gr-qc/0105121}.

\bibitem{NC} N. J. Nunes and E. J. Copeland, Phys. Rev. D{\bf 66}, 043524 (2002), \texttt{astro-ph/0204115}.

\bibitem{curvaton}A. R. Liddle and L. A. Urena-Lopez,
Phys. Rev. D{\bf 68}, 043517 (2003),
\texttt{astro-ph/0302054}.

\bibitem{portugas} M. C. Bento, R. G. Felipe, and N. M. C. Santos,
Phys. Rev. D{\bf 77}, 123512 (2008),
\texttt{arXiv:0801.3450 [astro-ph]}.

\bibitem{Komatsu5yrWMAP} E. Komatsu {\it et al.},
  Astrophys. J. Suppl.\  {\bf 180}, 330 (2009),
  \texttt{arXiv:0803.0547 [astro-ph]}.
 
\bibitem{Komatsu7yrWMAP} E. Komatsu {\it et al.}, 2010,
  \texttt{arXiv:1001.4538 [astro-ph]}.

\bibitem{wands}D. H. Lyth and D. Wands,
Phys. Lett. B{\bf 524}, 5 (2002),
\texttt{hep-th/0110002}.

\bibitem{curv2} S. Mollerach, Phys. Rev. D{\bf 42}, 313 (1990); K. Enqvist and M. S. Sloth, Nucl. Phys. B {\bf 626}, 395 (2002), \texttt{hep-ph/0109214}; T. Moroi and T. Takahashi, Phys. Lett. B{\bf 522}, 215 (2001), \texttt{hep-ph/0110096}.

\bibitem{curvreh} B. Feng and M. Li, Phys Lett. B{\bf 564}, 169 (2003), \texttt{hep-ph/0212213}.

\bibitem{anthony}A. R. Liddle and A. Smith,
Phys. Rev. D{\bf 68}, 061301 (2003),
\texttt{astro-ph/0307017}.

\bibitem{GW}D. Langlois, R. Maartens, and D. Wands,
Phys. Lett. B{\bf 489}, 259 (2000),
\texttt{hep-th/0006007}.

\bibitem{scaling}A. R. Liddle and R. Scherrer,
Phys. Rev. D{\bf 59}, 023509 (1999),
\texttt{astro-ph/9809272}.

\bibitem{SWZ} P. J. Steinhardt. L. Wang, and I. Zlatev, Phys. Rev. D{\bf 59}, 123504 (1999), \texttt{astro-ph/9812313}.

\bibitem{thawing}R. Caldwell and  E. Linder,
Phys. Rev. Lett. {\bf 95}, 141301 (2005),
\texttt{astro-ph/0505494}.

\bibitem{CLW} E.J. Copeland, A. R. Liddle, and D. Wands,
Phys. Rev. D{\bf 57}, 4686 (1998), \texttt{gr-qc/9711068}. 

\bibitem{LPB} A. R. Liddle, P. Parsons, and J. D. Barrow, Phys. Rev. D{\bf 50}, 7222 (1994), \texttt{astro-ph/9408015}.
\end{thebibliography}
\end{document}